# The contribution of the Joule-Thomson effect to solar coronal heating

Claudio Vita-Finzi
Dept of Earth Sciences, Natural History Museum
London SW7 5BD, UK  (cvitafinzi@aol.com)

23  December 2016

   Two of the three gases that display isenthalpic Joule-Thomson (J-T) warming under laboratory conditions are hydrogen and helium, the main constituents of the solar plasma, as the inversion point at which expansion results in heating is only ~51 K for helium and ~193 K for hydrogen, but the temperatures that are attained by this route are at most a few hundred K. Increases in ion temperature by several orders of magnitude are claimed for hydrogen plasmas subject to expansion into a vacuum; modest increases are reported for the shortlived tests of this effect that have been carried out in space in the wakes of artificial satellites and of the Moon. Attempts to calculate the J-T coefficient at very high temperatures using equations of state and thermodynamics  remain very preliminary. The  potential contribution of plasma expansion to heating of the solar corona must therefore be assessed empirically, but this is consistent with how the J-T effect was first identified. Observational data indicate an inverse relation between electron density ($n_e$) and temperature from the Sun's photosphere at least to 1 AU. The process is most effective at the transition zone between the chromosphere and the corona, where a steep rise in temperature from $10^3$ to $10^6$ K over a radial distance of ~ $10^3$ km is accompanied by a sharp fall in electron density. The sunspot record, EUV measurements by the EVE instrument on the SDO satellite, and solar wind fluctuations documented by the ACE satellite indicate broadly coherent periodicity from the photosphere to the outer corona consistent with a non-pulsatory heating process. It comprises three successive stages characterised by induction, the J-T mechanism, and plasma expansion. Astronomical data may therefore be used to derive rather than to test an extension of the J-T effect

*which could help to explain heating in other solar system bodies and other stellar coronae.*

The effect of pressure change on the temperature of real gases was first investigated in the laboratory in France by Gay-Lussac (1807) and in the UK by Joule (1845) and Thomson & Joule (1852). Consequent on these studies the Joule-Thomson (J-T) effect is a term now sometimes used to indicate warming of a real (i.e. not ideal) gas subject to free expansion under constant enthalpy if it rises above its inversion temperature ($T_i$) and cooling if it drops below its $T_i$. The corresponding J-T coefficient $\mu_{JT}$ denotes $(\delta T/\delta P)_H$ where T is the change of temperature, P is the decrease in pressure and H constant enthalpy. A subsidiary issue is the extent to which throttling – or, as Thomson & Joule (1852) put it, 'rushing through small apertures' – is critical to the effect: some sources apply the name Joule-Kelvin to expansion with throttling and the name Joule to free expansion (Goussard & Roulet 1993; Albarrán-Zavala et al. 2009).

The possible contribution of the J-T effect to coronal heating was recently broached (Vita-Finzi 2016a) on the flimsy grounds that the major constituents of the corona -- hydrogen and helium -- are two of the three gases that experience J-T warming (the third being neon) under laboratory conditions. Hydrogen and helium have unusually low $T_i$ s so that the inversion point at which $\mu_{JT}$ changes sign is low, namely ~51 K for helium and ~193 K for hydrogen. The question here is thus how these gases behave at pressures and temperatures well beyond normal experimental range and routine theoretical discussion so that the J-T effect can legitimately be invoked to account for coronal temperatures of up to and perhaps beyond $10^6$ K which remain largely unexplained.

Besides temperature and pressure our enquiry needs to address the tempo at which the effect operates, as some kind of continuous process is called for rather than the near-instantaneous response of the classic experiments and discussions: the candidate coronal gases are constantly being renewed whether because on the imbalance of the mechanical pressures between interplanetary space and the base of the solar corona or because heat conduction is not efficient enough to evacuate the excess energy deposited in the inner corona (Lemaire 2010).

**The J-T effect in the corona**

Published J-T diagrams imply that, as $\mu_{JT} > 0$ (and therefore cooling) is confined within the plotted inversion curve, heating (i.e. $\mu_{JT} < 0$) will prevail elsewhere, but the plots (e.g. Kent 1993) are generally limited to temperatures < 1000 K. Gold (1964) has shown that completely ionized gases subject to free expansion experience heating and also manifest an inversion temperature, but he cautioned (doubtless with MHD devices in mind) that the J-T effect will lead to 'thermonuclear' temperatures only 'in the densest possible plasmas': the maximum temperature change that emerges from his calculated expansion is ~ $\Delta P/P \times 10^5$ degrees, but in his view it will not be attained in practice both because the Debye length may be several orders of magnitude greater than the value used in the estimate (viz ~$10^{-8}$ cm) and because the heating process is subject to self-quenching.

There have been limited attempts to use equations of state and thermodynamics in order to consider the K-T effect at temperatures too high for routine direct measurement: for example, the second and third virial coefficients have been approximated for helium and used to calculate the zero-pressure J-T coefficient and its first pressure derivative (Mason & Vanderslice 1958). But appeals to the J-T effect to account for coronal properties require that we demonstrate a direct link between temperature and expansion, that any associated heating is cumulative, and that the process is effectively continuous and (unlike magnetic reconnection) neither spasmodic in incidence nor greatly variable in severity.

As things stand there is a circumstantial case for continuous though not necessarily uniform heating from the photosphere to the upper corona and beyond. Three mechanisms are involved: the generation of magnetic energy by rotating structures in the Sun's induction zone, heating of the hydrogen-alpha of the chromosphere by Foucault currents generated in the upper photosphere, and (as suggested here) J-T heating of the coronal gases. A preliminary test of this progression was made by plotting sunspot number (SSN) and average irradiance at 1 minute averages as 10-minute moving averages, over a 6 month period (Jan-June 2012, chosen for its anodyne character within the rising limb of Solar Cycle 24), for seven coronal lines monitored by the Extreme Ultraviolet Variability Experiment (EVE) instrument on the Solar Dynamics Observatory (SDO) satellite (Vita-Finzi 2016b). The chosen lines represent temperatures spanning

$10^3 - 10^{6.8}$ K, that is to say from the photosphere as manifested in sunspot number (SSN) through 304 Å, corresponding to the upper photosphere/transition zone, 193 Å at the upper corona, to 94 Å, which represents the corona during a solar flare (NASA 2013).

These identifications and temperature estimates are of course approximate. Thus the dominant contribution to the 94 Å channel observed by the Atmospheric Imaging Assembly (AIA) on the SDO comes from the Fe X 94.01 Å line ($10^{6.05}$ K) rather than Fe XVIII at 93.93 Å ($10^{6.85}$ K) (O'Dwyer et al. 2010). Even so the data span the three orders of magnitude of the problematic temperature ascent. The plots were broadly synchronous, with 7 major peaks clearly displayed throughout the array (Vita-Finzi 2016b). The fluctuations ranged from $10^{-4}$ W m$^{-2}$ for the 304 Å line to $10^{-6}$ W m$^{-2}$ for the 94 Å line, that is to say with reduced amplitude but consistent periodicity: prima facie evidence of the radial transmission of a controlling signal. Why the signal should fluctuate at source is unclear but the parallelism between the coronal and sunspot oscillations points to irregularities in the H/He flux as the primary modulating factor. The correspondence between EUV maxima and peaks in wind speed measured from the Advanced Composition Experiment (ACE) satellite (Fig. 1) - admittedly a coarse measure of coronal expansion - supports the notion of thermal continuity throughout the corona. The J-T stage, the most controversial, must now be evaluated.

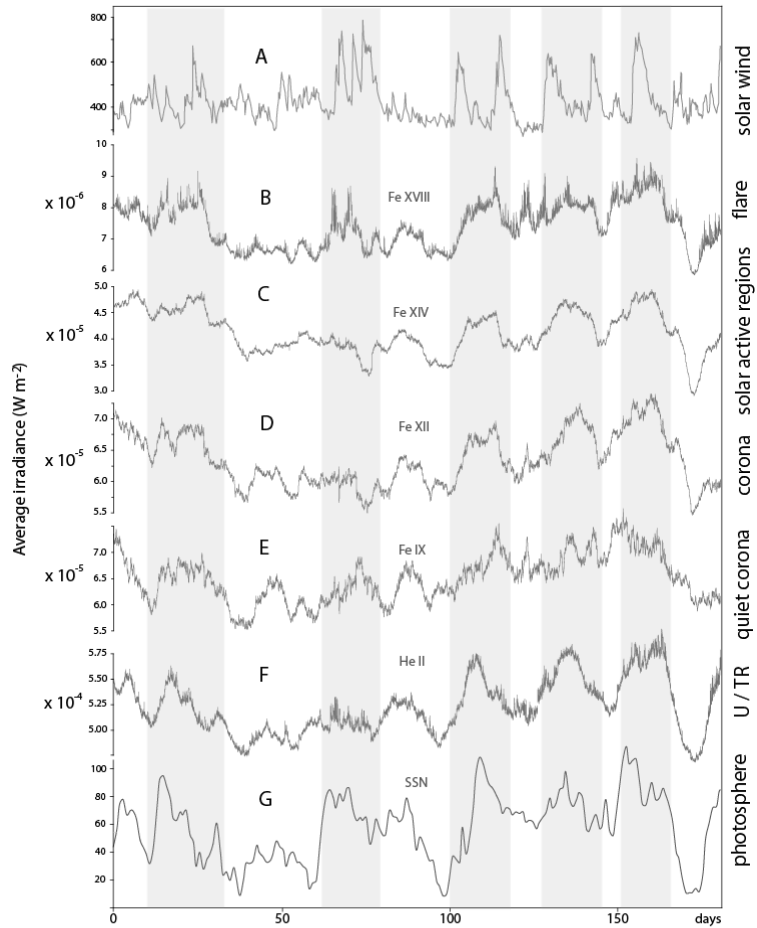

Fig 1 (A) solar wind speed, data courtesy of ACE/SWEPAM instrument team and ACE Science Center; (B-F) five coronal lines (EUV/SDO 1 min averages plotted as 10 min moving averages), data courtesy of lasp.colorado.edu; (G) net sunspot number, data courtesy of ngdc.noaa.gov, all for 1 Jan-1 July 2012. EUV (B-F) lines relation to solar atmosphere zonation (TR = transition region) based on Atmospheric Imaging Assembly (AIA/SDO) instrument wavelength bands (aia.lmsl.com) website. The lower 6 rows are based on Vita-Finzi 2016b, fig 8.3.

Theoretical studies and computer simulation have shown that the expansion of a plasma into a vacuum or a more tenuous plasma leads to the acceleration of ions greatly above their thermal energy. The effect was first demonstrated by Gurevich et al. (1966), who claimed that the acceleration could amount to several orders of magnitude and who referred to pertinent experimental studies which had also revealed ion acceleration above thermal levels. The bearing of this effect on space phenomena was made explicit by the interaction of an obstacle with a plasma. Samir & Wrenn (1972) reported that ionospheric electron temperature measured by a Langmuir probe in the near wake of an artificial satellite (Explorer 31) was raised above that of the ambient electron gas by as much as 50 %. They referred to earlier work (Medved 1969) on the Gemini/Agena spacecraft in which wake temperature was 1700 K greater than the ambient temperature in one experiment and 764 K in another. The Moon's wake provided scope for related work; the increase in the electron temperature in the lunar wake found by the SWE plasma instrument on the WIND spacecraft amounted to a factor of four although ion temperatures were little changed (Ogilvie et al. 1996). Laboratory investigations based on immersion of a plate in a single-ion, collisionless, streaming plasma, saw 'early time expansion' result in ion acceleration into the wake (Wright et al. 1985).

**Discussion**

The question remains whether any resulting heating is sustained and is cumulative so that temperatures of $10^6$ K can be attained. An inverse relation between pressure fall and temperature in an astronomical context was assumed by Kothari (1938) when he showed that, for a relativistically degenerate gas (i.e. one nearing its ground state) undergoing Joule-Thomson expansion, the degree of heating per unit fall of pressure increased with the degree of degeneracy, although he did not specify the range of this process. Industrial applications of J-T cooling from the outset amplified the J-T cooling by repeatedly recycling the gas via a heat exchanger. Our need is to explain heating in unidirectionally flowing plasma.

The outcome is at its most pronounced in the transition region (TR). Sandwiched between the Sun's chromosphere and its corona, the TR is a complex zone of spicules, fine structures and loops rather than a tidy layer of uniform thickness (Priest 2014). Even so it coincides with a net rise in temperature from ~3.5 x $10^4$ K to $10^6$ K over a radial distance variously put at $10^2$ -$10^3$ km. The corresponding change in gravitational scale height is therefore so slight that pressure across the TR is effectively unchanged, and density was consequently thought by Mariska (1986) to undergo a 40-fold reduction.

The estimated electron density for the TR as a whole is $10^{15}$ m$^{-3}$ (but see Buchlin & Vial 2009 for problems created by sharp density and temperature gradients) compared with $10^{19}$ m$^{-3}$ (Priest 2014) for the chromosphere, a cumulative fall rather more substantial than 1/40 but consistent with the progressive change in electron density (Fig. 2) represented by typical values of $10^{23}$ m$^{-3}$ for the photosphere, $10^{15}$ m$^{-3}$ for the TR as a whole, as we saw, $10^{14}$ for the base of the corona base in quiet regions, $10^{12}$ m$^{-3}$ at 1 R$_\odot$, $10^7$ m$^{-3}$ at 1 AU and $10^6$ m$^{-3}$ in the interstellar medium (Priest 2014, Aschwanden 2006). Empirical electron density profiles derived from white light brightness measurements (Lemaire 2012) range from $10^{15}$ m$^{-3}$ at 0.01 R$_\odot$ to $10^{10}$ m$^{-3}$ at 10 R$_\odot$ and indicate a similar gradient. Note that novel techniques already permit more subtle differentiation. Hard X-rays, for example, allow neutral gas density in the chromosphere to be distinguished with a resolution of ~150 km ; but the result is close to a linear decline in number density of over two orders of magnitude over a height of ~$10^3$ km above the photosphere (Kontar et al. 2008), by no means inconsistent with our electron plot.

The anomalously rapid rise in temperature and fall in pressure that define the TR support the case for a range of different mechanisms to account for coronal heating. We have previously suggested that EUV cyclones in the photosphere, modulated by convection in the solar subsurface, induce currents in the chromosphere thereby converting magnetic to thermal energy (Vita-Finzi 2016a). The chromospheric plasma is only partly ionised (Aschwanden 2009): the ionisation fraction of pure hydrogen rises steeply at about $10^4$ K, whereupon, as Gold (1964) noted, it becomes more susceptible to the J-T effect. This in turn is now supplanted once temperatures attain $10^6$ K by the process identified by Gurevich et al. (1966) and driven by plasma expansion. The three proposed stages are shown schematically in Figure 2.

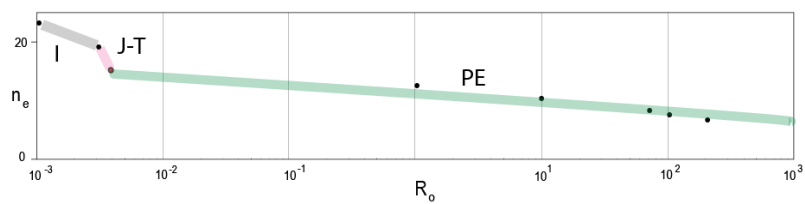

Fig 2   Electron density n$_e$ (m$^{-3}$) and radial distance (R$_\odot$) from photosphere. Data points derived from Aschwanden (2009), LeBlanc et al. (1998), Lemaire (2012) and Priest (2014). Range of proposed heating mechanisms: I (induction), J-T (Joule-Thomson), PE (plasma expansion).

Our focus on hydrogen and helium should not obscure the potential of the proposed mechanism for other solar system gases. The inversion temperature for $CH_4$ for instance is 1290 K and for $CO_2$ it is 2050 K. The proposed scheme could thus help to explain heating in other bodies (such as Titan) which display a radial increase in temperature, a decrease in plasma density, and sustained gas outflow. It may also bear on the thermal evolution of other coronal stars.